\documentclass[12pt]{article}

\newcommand{\mnras}{MNRAS}
\newcommand{\apj}{ApJ}
\newcommand{\apjl}{ApJ}
\newcommand{\nat}{Nature}
\newcommand{\apjs}{ApJS}
\newcommand{\aap}{A\&A}
\newcommand{\araa}{ARA\&A}
 
\newcommand{\aj}{AJ}

\usepackage{scicite}

\usepackage{times}
\usepackage{graphicx}
\usepackage{lineno}
\usepackage{setspace}
\usepackage{caption}
\usepackage{amsmath}
\usepackage{amssymb}

\def\msun{M$_\odot$}

\newenvironment{sciabstract}{%
\begin{quote} \bf}
{\end{quote}}

\newcounter{lastnote}

\title{A single population of red globular clusters around the massive compact galaxy NGC 1277}

\author
    {Michael A. Beasley,$^{1,2*}$ Ignacio Trujillo,$^{1,2}$ Ryan Leaman,$^{3}$ Mireia Montes,$^{4}$\\
\normalsize{$^{1}$Instituto de Astrof\'isica de Canarias, Calle V\'ia L\'actea, La Laguna, Tenerife, Spain}\\
\normalsize{$^{2}$University of La Laguna. Avda. Astrof\'isico Fco. S\'anchez, La Laguna, Tenerife, Spain}\\
\normalsize{$^{3}$Max-Planck Institut f\"ur Astronomie, K\"onigstuhl 17, D-69117, Heidelberg, Germany}\\
\normalsize{$^{4}$Department of Astronomy, Yale University, 06511 New Haven, CT, USA}\\
}

\date{}

\begin{document} 




\maketitle


\begin{sciabstract}
  
Massive galaxies are thought to form in two phases: an initial, early collapse of gas and giant burst
of central star formation, followed by the later accretion of material that builds up their
stellar and dark matter haloes\cite{Kochfar2006}\cite{Oser2012}\cite{Ceverino2015}\cite{vanDokkum2010}.
The globular cluster systems of such galaxies are believed to form in a similar manner.
The initial central burst forms metal-rich (red) clusters, while more metal-poor
(blue) clusters are brought in by the later accretion of less massive
satellites\cite{SearleZinn1978}\cite{Cote1998}\cite{Beasley2002}\cite{Tonini2013}\cite{Leaman2013}\cite{Kruijssen2015}.
This formation process is thought to lead the creation of the multimodal optical colour distributions
seen in the globular cluster systems of massive galaxies \cite{Tonini2013}\cite{Peng2006}\cite{BrodieStrader2006}.
Here we report optical observations of the massive relic galaxy NGC~1277, 
a nearby unevolved example of a high redshift ``red nugget" \cite{Damjanov2009}\cite{vandenBosch2012}\cite{Trujillo2014}
\cite{MartinNavarro2015}\cite{Yildirim2017}.
The $g_{\rm 475W}-z_{\rm 850LP}$  cluster colour distribution shows that the 
GC system of the relic is unimodal and uniquely red.
This is in strong contrast to normal galaxies of 
similar and larger stellar mass, whose cluster systems always exhibit (and are generally dominated by)
blue clusters\cite{Peng2006}. We argue that the cluster system of NGC~1277 indicates that the
galaxy has undergone little (if any) mass accretion after its initial collapse and 
use analytic merger trees to show that the total stellar mass accretion is likely less than $\sim10$\%. 
These results confirm that NGC~1277 is a genuine relic galaxy and show
that the blue clusters constitute an accreted population in present day massive galaxies.

\end{sciabstract}


\section*{Introduction}

We obtained HST/ACS $g_{\rm 475W}$ and $z_{\rm 850LP}$ imaging of NGC~1277 in order to characterize its globular cluster
system (E.D. Fig.~1).
NGC~1277 has been identified as a massive ($\sim1.2\times10^{11}$~\msun; \cite{Trujillo2014})
relic galaxy based on its high stellar mass density, compactness, kinematics and old and metal-rich stellar
populations \cite{vandenBosch2012}\cite{Trujillo2014}\cite{MartinNavarro2015}\cite{Yildirim2017}.
We hypothesized that if NGC~1277 is a true relic -- a low redshift counterpart of massive, compact high-redshift
galaxies termed ``red nuggets'' -- then NGC~1277 should have accreted little or no stellar
halo and should have few or no blue clusters.

The spatial distributions of cluster candidates from the HST/ACS imaging (Fig.~1)
indicate that both NGC~1277 and its companion galaxy in projection, NGC~1278, have relatively rich cluster systems.
Separating the clusters in the ACS field by colour by cutting at ($g_{\rm 475W}-z_{\rm 850LP})_0 = 1.15$,
the approximate separation between the red and blue peaks for a galaxy of this mass in the ACS Virgo
Cluster Survey\cite{Peng2006}, it is apparent that while NGC~1277 has a rich red cluster population,
it has few blue clusters. This is in strong contrast to NGC~1278, which has a similar
stellar mass to NGC~1277 ($\sim2.4\times10^{11}$~\msun), but has both red and blue clusters.

This visual impression is supported by looking at the detailed colour distribution of the NGC~1277
GCs compared to galaxies from the Virgo Cluster survey of similar stellar mass (Fig.~2).
No obvious blue peak is seen in the cluster colour distribution of NGC~1277, which appears red
with a tail to bluer colours. In contrast, the survey composite colour distribution has both blue and red peaks.
We ran gaussian mixture modelling tests ({\sc gmm}) (Methods) on these colour distributions
whose null hypothesis is that the input colour distributions are unimodal.
For NGC~1277,  {\sc gmm} cannot reject the null hypothesis, i.e. that of  a unimodal distribution for the
GCs ($p(\chi^2) = 0.371$).
However, under the assumption that two populations of clusters are indeed present, {\sc gmm} associates 99 clusters
with a red cluster population with a mean colour ($g_{\rm 475W}-z_{\rm 850LP})_0 = 1.31\pm0.10$
while 21 clusters are associated with a metal-poor component with mean colour $(g_{\rm 475W}-z_{\rm 850LP})_0 = 0.86\pm0.24$.
For the survey colour distribution, {\sc gmm} rejects a unimodal distribution at
high confidence ($p(\chi^2)<0.001$), finding a roughly even split between blue (101 clusters) and red clusters (110 clusters) with colour
peaks at ($g_{\rm 475W}$-$z_{\rm 850LP})_0 = 0.97\pm0.04$ and $1.36\pm0.04$, respectively.

The results of the survey show that the cluster systems of present-day massive galaxies can have
diverse colour distributions and that all galaxies in the survey, independent of their stellar mass, have
substantial numbers of blue clusters \cite{Peng2006}.
For galaxies of stellar masses comparable to NGC~1277, the typical blue globular
cluster fraction is $0.50\pm0.10$ \cite{Peng2006} or larger.
For NGC~1277 we find a blue cluster fraction of, $f_{blue} = 0.17^{+0.12}_{-0.17}$ (systematic)$^{+0.06}_{-0.05}$ (random)
  (Fig.~3, Methods).
This is an upper limit since the null hypothesis of unimodality is not rejected.
NGC~1277 is clearly an outlier in that it has a significantly lower blue fraction
than the survey galaxies at similar stellar mass (Fig.~3).

For massive galaxies in the survey, the blue cluster fractions are lower limits because the survey only covered the
inner 1-2 effective radii ($R_e$) of these galaxies, which is a region where red clusters
dominate over blue clusters due to the red clusters more concentrated spatial distributions~\cite{Hargis2014}.
In the case of NGC~1277, which is very compact ($R_e = 1.2$ kpc) and located in the more distant Perseus cluster,
we have probed the entire radial range ($\sim10~R_e$)
of the cluster system and therefore are not biased against detecting blue clusters.
Thus, NGC~1277 truly lacks blue clusters when compared to normal galaxies (Fig.~3).

We rule out the possibility that a colour-magnitude relation in the blue
GCs is driving the red-dominated peak of the NGC~1277 clusters.
This is not seen in our data (Methods, E.D. Fig.~6).
Our results are also unaffected by possible non-linear colour-metallicity
  relations of the clusters \cite{Blakeslee2012} \cite{Yoon2011}\cite{ChiesSantos2012}.
The majority of clusters ($\sim$70\%) in NGC~1277 have $(g_{\rm 475W}-z_{\rm 850LP})_0 > 1.1$, which
is the approximate colour where the empirical $(g_{\rm 475W}-z_{\rm 850LP})_0$ -- metallicity
relations for clusters in the Milky Way, M49 and M87 are linear\cite{Peng2006}.

  NGC~1277 is $\sim5$ times more compact than normal galaxies of the same stellar mass at $z = 0$~\cite{Trujillo2014}.
As we have shown, NGC~1277 also lacks a significant population of blue clusters which are generally
associated with a ``halo'' population in galaxies. The lack of a halo is
  consistent with the surface brightness profile of NGC~1277 which shows a sharp
  decline beyond $\sim10$ kpc corresponding to a factor of $\sim5$ less stars
  when compared to normal galaxies of the same mass~\cite{Trujillo2014}.
Assuming that NGC~1277 obeyed the same size-stellar mass relation
for quiescent galaxies of the same stellar mass at $z = 2$, NGC~1277 has either been stripped of
its outer envelope and clusters and been reduced in size, or NGC~1277 has failed to grow substantially in size and
its blue cluster system since $z = 2$. We explored both scenarios and favour the latter
picture -- i.e. NGC~1277 is a genuine relic galaxy.

The tidal radius of NGC~1277 -- the radius at which stars and clusters become
unbound from the galaxy -- can be estimated analytically \cite{Read2006}.
In the extreme case that NGC~1277 is currently at pericentre within the
Perseus cluster and that its stellar halo and clusters are on prograde orbits with respect to its orbit
within Perseus, we calculate a lower limit for the tidal radius of 11 kpc. This corresponds to
$\sim2-3$ effective radii in a normal sized galaxy of NGC~1277's stellar mass and is a radius where
the cluster population should
consist of at least 50\% blue clusters \cite{Peng2006}. This suggests that stripping has not significantly
affected the physical size or the cluster system of NGC~1277.
Furthermore, the central stellar density of NGC~1277
is a factor of $2-3$ higher than normal galaxies within 1 kpc~\cite{Trujillo2014}.
Tidal stripping cannot increase central stellar densities \cite{Read2006} implying that
NGC~1277 was born both dense and compact. This then raises the question of why NGC~1277 should
suffer severe tidal stripping while larger galaxies of similar mass in the same
environment -- such as NGC~1278 -- have not.
Finally,  the outer region of NGC~1277 shows no evidence
for tidal tails or streams \cite{Trujillo2014} and its kinematics show
no evidence for stripping in its stellar orbits, in its rotation curve or
mass distribution \cite{Yildirim2017}.

In contrast, NGC~1277 strongly resembles the low-redshift equivalent of higher redshift compact systems.
NGC~1277 lies on the size-stellar mass relation of $z = 2$ galaxies \cite{Trujillo2014} \cite{Yildirim2017}.
It has a high central stellar density consistent with massive galaxies
at these redshifts \cite{Szomoru2012}. Schwarzschild dynamical modelling 
\cite{Yildirim2015} shows that NGC~1277 is dominated by high angular momentum orbits
implying that it formed in a largely dissipative process. The results of \cite{Yildirim2017}\cite{Yildirim2015}
imply that NGC~1277 is essentially a massive disk with no classical, pressure-supported component.
These kinematical properties are remarkably similar to the lensed ``dead disk'' at $z\sim2.1$\cite{Toft2017},
which has the same stellar mass as NGC~1277 and is similarly compact ($R_e = 1.73^{+0.34}_{-0.27}$ kpc).
Additionally, the old, metal-rich stellar population of NGC~1277 is similar to other nearby
relic-galaxy candidates \cite{Yildirim2017}, with the expectations for passively evolving  $z = 2$ red nuggets and with
predictions for the redshift evolution of the massive, compact systems in cosmological simulations~\cite{Wellons2016}\cite{Furlong2017}. Finally, the fact that NGC~1277 lies in a massive galaxy cluster is neither unexpected observationally or theoretically (Methods).

Our findings for the cluster system of NGC~1277 bring further contraints on the evolutionary histories
of relic galaxies.
In accretion models for the formation of the blue cluster population in massive galaxies, the location of the blue peak
and the relative fraction of blue clusters brings insight into the total mass and properties of accreted satellites by any
given galaxy\cite{Cote1998}\cite{Tonini2013}\cite{Leaman2013}.
We ran a series of analytic merger models (Fig.~4, Methods, E.D. Figs.~7--8) in order to constrain the 
accreted mass fraction of NGC~1277, based on our observations.
From this modelling we find that the observed blue cluster fraction of NGC~1277 is characteristic of galaxies
which have accreted $\sim10$\% or less of their stellar mass, with the rest of the stars forming {\it in-situ}.
By contrast,  mass-matched ACS Virgo cluster survey galaxies show accretion fractions of $50-90$\%.
Therefore, we find that the blue fraction of globular clusters approximately traces the total fraction of accreted mass
in galaxies, and that NGC~1277 has accreted very little mass (if any) over its lifetime in comparison with other galaxies
of similar stellar mass.

Hence we conclude that NGC~1277 provides a nearby benchmark for the detailed study of the properties
of massive galaxies which remain unevolved since $z = 2$. High-resolution imaging of the cluster systems of more
relic galaxies will allow for the reconstruction of their mass accretion histories.

\newpage

\begin{figure*}
    \begin{center}
\includegraphics[scale=0.4]{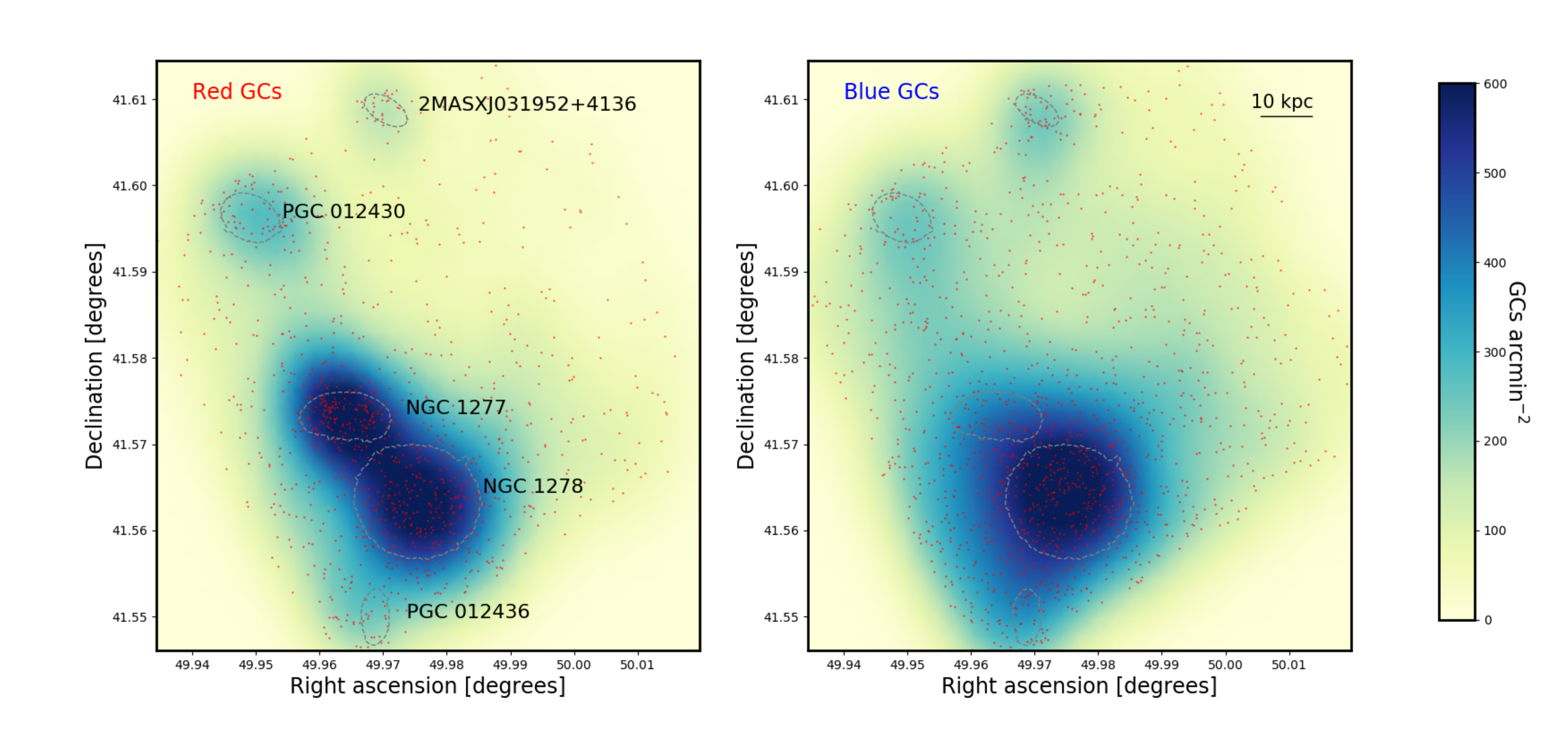}
\caption*{{\bf Fig. 1. Spatial distribution of clusters in HST/ACS field.} North is up, East is to the right
  here. Individual clusters are shown as points.
  The locations of galaxies are indicated by galaxy isophotes corresponding to 23 mag arcsec$^{-2}$
  in $g_{\rm 475W}$.
  NGC~1277 is to left of centre in the plots, with the neighbouring galaxy 
  NGC~1278 located some 50 arcseconds ($\sim17$ kpc) to the South East in projection.
  The red and blue clusters have been separated by taking a cut at ($g_{\rm 475W}-z_{\rm 850LP})_0 = 1.15$,
  typical of the peak separation between the red and blue clusters in galaxies of this stellar mass\cite{Peng2006}.
  Overplotted is a gaussian kernel density estimate map constructed with a kernel of width
    10 arcseconds.
  Galaxies in the field can clearly be identified by the relative prevalence of clusters. NGC~1278
  shows both red and blue cluster populations, whereas NGC~1277 only has a significant population of red clusters.
}\label{Fig1}
\end{center}
\end{figure*}

\newpage

\begin{figure*}
  \begin{center}
\includegraphics[scale=0.6]{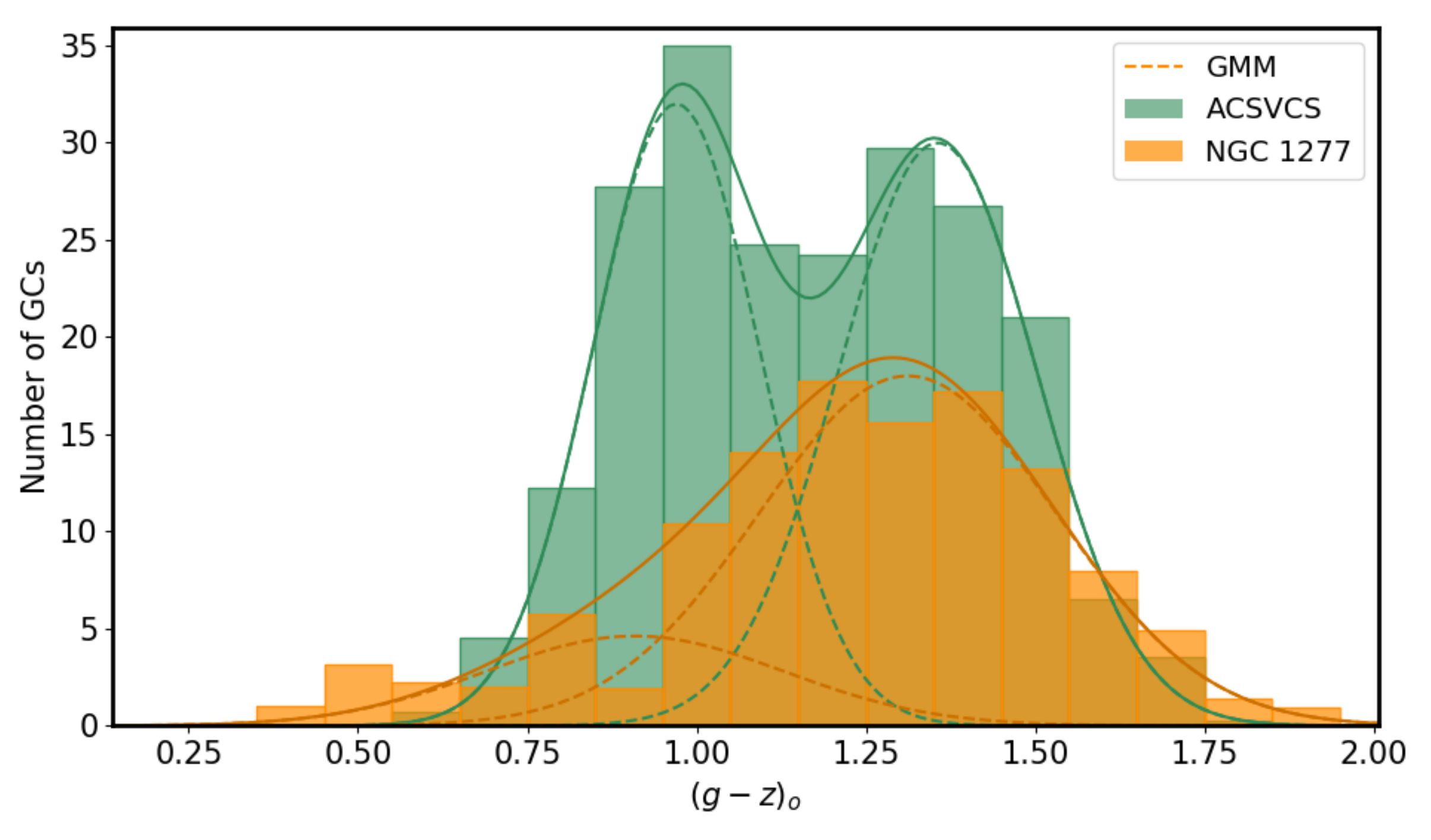}
\caption*{{\bf Fig. 2. The colour distribution of clusters in NGC~1277 compared to composite cluster system.}
  The colour distribution for NGC~1277 has been constructed for the entire radial extent
  of its cluster system (to 11 kpc), and has been corrected for the contamination from NGC~1278
  clusters which, in projection, is 17 kpc from NGC~1277. The composite cluster system
  was constructed from four galaxies from \cite{Peng2006}
  with similar stellar masses to NGC~1277 ($M_*\sim1.2\times10^{11}$\msun). The dashed
  curves indicated Gaussian components of the cluster systems obtained from gaussian mixture modelling
  with {\sc gmm} \cite{Muratov2010}. The composite cluster system is clearly bimodal, while NGC~1277 lacks an obvious
  blue peak.
}\label{Fig2}
\end{center}
\end{figure*}

\begin{figure*}
    \begin{center}
\includegraphics[scale=0.7]{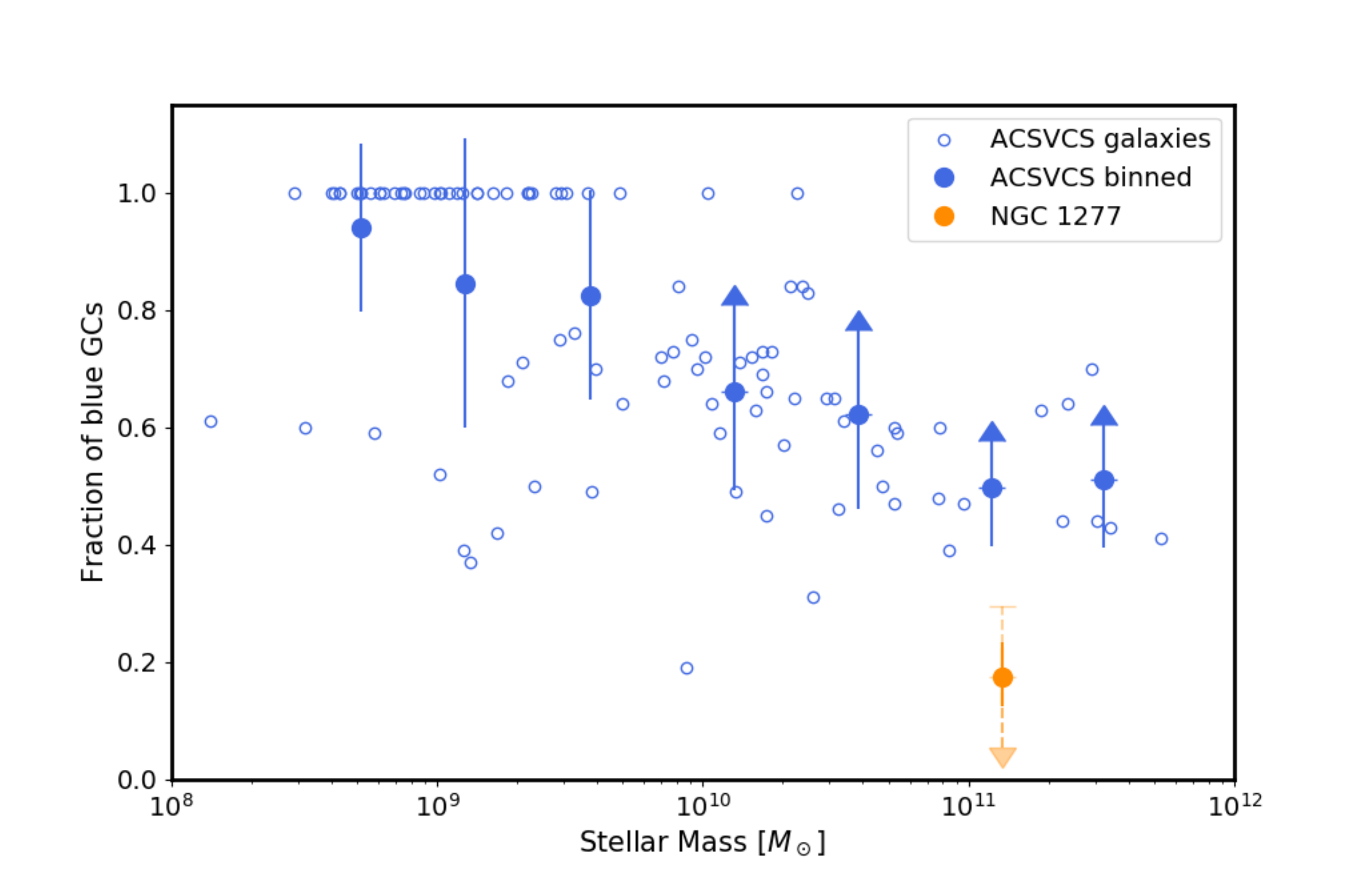}
\caption*{{\bf Fig. 3. The fraction of blue clusters in galaxies of a given stellar mass.}
  Data for the survey galaxies comes from \cite{Peng2006}.
  The figure shows that NGC 1277 has a very small (or zero) fraction of blue clusters
  when compared to normal sized galaxies of the same mass.
  The blue fractions for the brightest survey galaxies are lower limits (indicated in the
  binned data showing 16, 84 percentiles) since only the central parts of the galaxy cluster systems were observed.
  On the contrary, the blue fraction for the
  NGC~1277 clusters is an upper limit since the null hypothesis (i.e., the complete absence of a gaussian
  distribution of blue clusters) cannot be rejected observationally.  The systematic uncertainty
  on the blue fraction of NGC~1277 is shown as the faint, dashed vertical errorbar (16, 84 percentiles),
  the random uncertainties from Monte Carlo simulations are smaller, solid vertical lines (16, 84 percentiles).
}\label{Fig3}
\end{center}
\end{figure*}

\begin{figure*}
    \begin{center}
\includegraphics[scale=0.7]{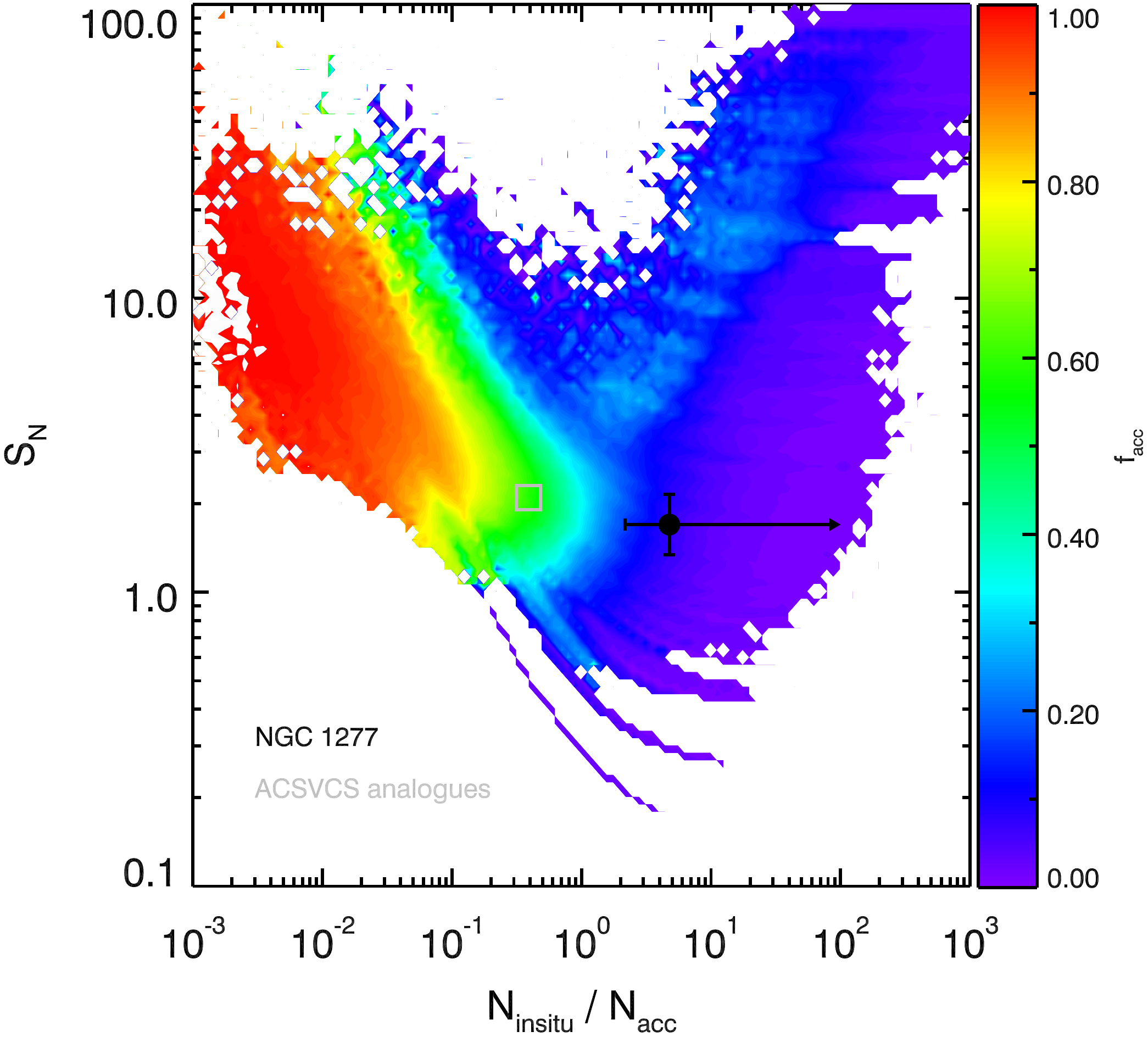}
\caption*{{\bf Fig. 4: Total specific frequency ($S_{\rm N}$) vs. number ratio of in-situ to accreted clusters
    in the simulated merger histories of NGC~1277.} The colours represent the average accretion fraction
  of merger histories that fall in a particular region of this parameter space.  Observed constraints for
  NGC 1277 (black point) are satisfied regularly in low accretion fraction ($\sim 10\%$) histories. Error bars
  are 16, 84 percentiles of the distribution.
  Higher accretion fraction merger histories rarely reach the observed colour ratio, without overproducing
  the total number of clusters.  In contrast, similar mass normal galaxies in the
ACS survey are reproduced well by merger histories with $f_{acc} \sim 50\%$ (grey box).
}\label{Fig4}
\end{center}
\end{figure*}

\section*{References}

\begingroup
\renewcommand{\section}[2]{}%

\endgroup

\subsection*{Acknowledgements}

We thank Claudio dalla Vecchia, Jorge Sanchez Almeida, Chris Brook, Sarah Wellons, Morgan Fouesneau, Alejandro Vazdekis,
Bililign Dullo, Justin Read and Glenn van de Ven for useful discussions and Javier Roman and Alejandro Serrano Borlaf for
assistance with image alignment.
M.A.B. and I.T.  acknowledge support from grant AYA2016-77237-C3-1-P from the Spanish Ministry of Economy and
Competitiveness (MINECO).
R.L. acknowledges funding from the Natural Sciences and Engineering Research Council of Canada PDF award.
This research has made use of the NASA/IPAC Extragalactic Database (NED) which is operated by the Jet Propulsion
Laboratory, California Institute of Technology, under contract with the National Aeronautics and Space Administration.
Based on observations made with the NASA/ESA Hubble Space Telescope, which is operated by the Association of
Universities for Research in Astronomy,
Inc., under NASA contract NAS 5-26555. These observations are associated with program GO-14215.
Support for this work was provided by NASA through grant HST-GO-14215 from the Space Telescope Science Institute,
operated by AURA, Inc. under NASA contract NAS 5-26555 This research has made use of NASA's Astrophysics Data System.
This work made extensive use of Python and Scipy.\\

\subsection*{Author contributions}

M.A.B. led the data processing and analysis, contributed to the interpretation and HST proposal preparation
and produced Figs 1--3, ED Figs 2--6.
I.T. contributed to the analysis and the interpretation, produced E.D Fig. 1 and lead the HST proposal preparation.
R.L. generated and analysed the analytic merger models, produced Fig. 4, E.D. Figs 7--8, contributed to the analysis,
interpretation and HST proposal preparation. M.M. contributed to the analysis, the interpretation and HST proposal preparation.
All authors contributed to the overall design of this project.\\

\subsection*{Author information}

The authors declare no competing financial interests. Correspondence and requests for materials should be
addressed to M.A.B. (beasley@iac.es)
  
\section*{Methods}
\subsection*{HST/ACS Photometry}

We obtained HST/ACS imaging of NGC~1277 in the F475W ($g_{\rm 475W}$) and F850LP ($z_{\rm 850LP}$)
(2 orbits; GO: 14215; PI: Trujillo)(E.D. Fig.~1). The total exposure time  was: 2280s (F475W)
and 2432s (F850LP).
We adopt a standard cosmological model with
$H_0$ = 70 km s$^{-1}$ Mpc$^{-1}$, $\Omega_m$ = 0.3 and $\Omega_\Lambda$ = 0.7.
The redshift assumed here for NGC~1277, z = 0.0169, corresponds to a galaxy distance of
73.3 Mpc ($(m-M)_0 = $34.33) and a spatial scale of 344 pc/arcsec.

The HST data were pipeline-reduced including correction for charge-transfer efficiency.
We calculated zeropoints for the AB magnitude system using the file header information.
Aperture and PSF photometry was performed on the ACS imaging using {\sc source extractor} (SE; \cite{Bertin1996})
and the SE add-on {\sc psfex} \cite{Bertin2011}. Our photometry was corrected for foreground extinction
using line of sight reddening estimates\cite{Schlafly2011}.
SE was run in dual image mode using unsharp masks as detection images with photometry performed on the
original images. We selected matched sources in the $g_{\rm 475W}$ and $z_{\rm 850LP}$ filters.
Aperture magnitudes were measured using a range of aperture radii from 3 to 50 pixels. We
applied aperture corrections to the 3-pixel aperture magnitudes by correcting to 10-pixel 
(0.5 arcsec) apertures based on bright, isolated stars, and then correcting to infinity using the corrections
tabulated in \cite{Bohlin2016}.
For artificial point-source tests, we used {\sc psfex} to construct PSFs based on bright, isolated stars
across the ACS field.

  \subsection*{Completeness tests}
    
Completeness tests were performed by injecting artificial point-sources into the images (using the
{\sc psfex} PSFs) and determining the recovery fraction as a function of magnitude with SE.
In regions $\sim20$ arcsec from bright galaxies we are 100\% complete to $g_{\rm 475W} = 27.3$ mag,
$z_{\rm 850LP} = 26.7$ mag with typical colour uncertainties of $\delta (g_{\rm 475W}-z_{\rm 850LP}) = 0.08$
mag. We take this as a conservative photometric limiting magnitude.
However, the completeness varies as a function of position and therefore we also calculated
photometric completeness by repeating our tests as a function of position and magnitude  across the field.
In the very central regions ($<$1.5 arcsec; 0.5 kpc) of NGC~1277~and neighbouring NGC~1278
we recover few clusters due to the high surface brightness of the galaxies.
Photometric errors were characterised by measuring the difference between the input and output  magnitudes of
50,000 artificial point sources placed randomly across the images.
To separate point sources (GC candidates) from extended sources we obtained
the SE output {\textsc FLUX\_RADIUS} as a function of magnitude for our artificial point sources. All
real sources identified within the region defined by the artificial sources were regarded
as cluster candidates.
In total, to $g_{\rm 475W} = $27.3 mag, we detect 2286 objects in the magnitude range consistent with being clusters.
Colour-magnitude diagrams (CMDs) for all the cluster candidates, and cluster candidates
around NGC~1278, NGC~1277 and an example background field are shown in E.D. Fig. 2.

\subsection*{Characterising the cluster system colour distributions}

The spatial density distribution of candidate clusters (Fig.~1) was constructed directly
from the photometry, after masking a small region (3 arcsecond radius) centred on an
uncatalogued dwarf located $\sim16$ arcseconds to the west of NGC~1277 (RA: 03:19:50.1, Dec: +41:34:22.10).
A kernel density estimate map was constructed using the python SciPy routine
{\sc gaussian\_kde} with gaussian kernel determined by ``Scott's rule'' \cite{Scott1992}.
As expected, the globular clusters aggregate around known galaxies (Fig.~1). However, taking a colour cut at
($g_{\rm 475W}-z_{\rm 850LP})_0 = 1.15$, the approximate separation between the red and blue peaks for a galaxy
the mass of NGC~1277 in ~\cite{Peng2006}, NGC~1277 seems to ``drop out'' of the
blue cluster density map. This contrasts with NGC~1278 that has both red and blue clusters clustered about the galaxy.

To create the colour distribution of NGC~1277 (Fig.~2), we were required to deal with interloping clusters from
the neighbouring NGC~1278 and also contributions from intra-cluster clusters.
We constructed a ``master background'' by selecting five regions located at
a distance from NGC~1278 equal to the separation between NGC~1277 and NGC~1278, and sufficiently distant from
NGC~1277 so as not to overlap with its cluster system (which has extent $\sim11$ kpc; E.D. Fig.~3).
The colour distribution of NGC~1277 clusters was built by selecting all cluster candidates within 11 kpc
($\sim10$ galaxy $R_e$) and subtracting off the master background normalised to the area of that covered
by the cluster system of  NGC~1277. This process also removes intra-cluster clusters
from the cluster system of NGC~1277. The colour distribution of this master background is shown
in E.D. Fig.~4.
The composite cluster system in Fig.~2 was constructed by selecting galaxies
from ~\cite{Peng2006} that bracket the stellar mass of NGC~1277, for a mean stellar mass,
$M_*\sim1.1\times10^{11}$~\msun~ with dispersion $0.4\times10^{11}$~\msun. To compare
with NGC~1277, we took a magnitude cut in the survey data at $g_{\rm 475W} = 23.98$ (Virgo cluster
distance modulus $(m-M)_0 = $31.01), which corresponds to our photometric depth.

To explore the sub-populations in the colour distributions, we ran a gaussian
mixture modelling code {\sc gmm} \cite{Muratov2010} on the binned colour distributions.
For the visibly bimodal survey composite colour distribution,
in the heteroscedastic (different variances between populations) case we obtain means of
($g_{\rm 475W}-z_{\rm 850LP})_0 = 0.97\pm0.04$ and $1.36\pm0.04$ with gaussian full-width half-maxima (FWHM)
of $0.12\pm0.02$ and $0.15\pm0.02$. 101 clusters are associated with the blue peak
and 110 clusters are placed in the red peak. A unimodal distribution is rejected at
high confidence ($p(\chi^2)<0.001$). Similar results were obtained in the
homoscedastic case.

For NGC~1277, a unimodal distribution cannot be rejected as the null hypothesis has
$p(\chi^2) = 0.371$ confidence.
Notwithstanding the possibility of unimodality, in the homoscedastic case for two populations,  we obtain means of
($g_{\rm 475W}-z_{\rm 850LP})_0 = 0.86\pm0.24$ and $1.31\pm0.10$ with FWHM
of $0.21\pm0.08$. 21 clusters are placed in the blue peak and 99 clusters are placed in the red peak (Fig.~2).
We define a blue fraction, $f_{blue} = N_{blue} / (N_{blue} +  N_{red})$.

  We consider two sources of uncertainty for $f_{blue}$. The uncertainty
  returned from {\sc gmm} on the number of blue and red clusters we regard as
  our systematic uncertainty. In addition, a random uncertainty on $f_{blue}$ comes from our
  background subtraction. In order quantify the true value of $f_{blue}$ and this uncertainty,
  we performed 1,000 Monte Carlo simulations where we randomly selected clusters from our background
  regions (until the observed background level was reached), and varied the radial apertures of the background
  and NGC~1277 selection
regions by $\pm$3 kpc ($\pm9$ arcseconds). From this we built new colour distributions and ran
{\sc gmm} on these colour distributions in order to obtain $f_{blue}$ (E.D. Fig.~5).
In the homoscedastic case, we obtain $f_{blue} = 0.18^{+0.12}_{-0.17}$ (systematic) $\pm0.04$ (random),
for the heteroscedastic case $f_{blue} = 0.17^{+0.12}_{-0.17}$ (systematic)$^{+0.06}_{-0.05}$ (random).
We conservatively take this latter value as the true blue fraction
and its associated uncertainties. From the Monte Carlo simulations, we find
$N_{blue} / N_{red} = 0.21^{+0.25}_{-0.20}$ (systematic)$\pm0.05$ (random).
We consider these are upper limits since the unimodal hypothesis cannot be rejected.

As a sanity check, we compared the colour distributions of NGC~1277 with NGC~1278 (E.D. Fig~4).
We remove contributions from intra-cluster clusters to the colour distribution
of NGC~1278 by selecting clusters in a region away from bright galaxies
(centred on RA:03:19:57.0, Dec: +41:35:30.0) over an area  normalised to that of our cluster selection region.
NGC~1278 shows a prominent blue peak, and an
evident though less prominent red peak.  {\sc gmm} (heteroscedastic case) locates peaks at
($g_{\rm 475W}-z_{\rm 850LP})_0 = 0.95\pm0.24$ and $1.40\pm0.10$. These solutions
are in excellent agreement with expectations \cite{Peng2006}.

\subsection*{Surface density profiles of the clusters}

The surface density profile of clusters (E.D. Fig.~3) was constructed by counting
GCs in semi-circular annuli and dividing by the area of each semi-annulus. We 
only counted clusters in the northern half of the galaxy -- bisected by the galaxy major axis --
to minimise contamination from the cluster system of NGC~1278.
The inner two radial bins have been corrected based on our
completeness tests. The clusters closely follow the galaxy light\cite{Trujillo2014}.
This is a characteristic property of red clusters in massive galaxies\cite{Pota2013}.
In contrast, blue clusters generally have spatial distributions that
are more extended than the galaxy stars \cite{Pota2013}\cite{Hargis2014}.
To define the radial extent of the system, and locate the background level,
we fit the data with a modified  S\'ersic function \cite{Sersic1968}:

\begin{equation}
  N_{GC}(R) = N_{e}~\times~exp\Bigg(-b_n\Bigg[\Bigg(\frac{R}{R_e}\Bigg)^{1/n} - 1 \Bigg]\Bigg)+bkg
\end{equation}

with $N_e$ being the surface density of clusters at radius $R_e$, $n$ is the S\'ersic index
and $bkg$ is the background value. $b_n$ is linked to $n$ as $b_N = 1.9992n - 0.3271$.

A S\'ersic fit ($N_e = 7.8\pm0.7$ clusters kpc$^{-2}$,  $R_e = 2.6\pm0.2$ kpc, $n = 0.9\pm0.3$, bkg = $0.27\pm0.01$ clusters kpc$^{-2}$)
to the surface density profile of the clusters (E.D. Fig.~3), and visual inspection, indicates that we reach
the background level at $\sim11$ kpc from the centre of NGC~1277. We consider this the full radial extent of the
NGC~1277 cluster system.\\

\subsection*{GC luminosity functions}

We constructed cluster luminosity functions for the NGC~1277 clusters in order
to obtain an independent distance estimate to the galaxy using the turn over of the cluster
luminosity function as a standard candle
\cite{Ferrarese2000}, and also to calculate the total size of the cluster system.
Again, we only selected from the northern half of the galaxy in
order to minimise interlopers from NGC~1278.
We binned the  $g_{\rm 475W}$ and 
$z_{\rm 850LP}$ magnitudes of the clusters as a function of magnitude (0.5 mag bins) and 
ran a version of the maximum likelihood code of  \cite{Secker1993} which fits for the mean ($\mu$), 
the full-width half-maximum ($\sigma$) and normalisation of the distribution.
The  code takes into full account the background, incompleteness and photometric errors.
In $g_{\rm 475W}$ ($z_{\rm 850LP}$) we obtain $\mu$ = $26.95\pm0.15$ mag ($25.80\pm0.20$ mag) and
$\sigma$ = $1.25\pm0.15$ ($1.35\pm0.20$) respectively. Assuming ``universal'' absolute magnitudes for the turn-over of the
luminosity function of $Mg_{\rm 475W} = -7.2$ mag ($Mz_{\rm 850LP} = -8.4$ mag) \cite{Jordan2007}, we obtain distance moduli
of $(m-M)_0 = 34.15\pm0.15$ ($(m-M)_0 = 34.20\pm0.20$). These values are  in excellent agreement with
our adopted distance to NGC~1277.\\

\subsection*{Total cluster population of NGC~1277}

We counted the total number of clusters brighter than the cluster luminosity function turnover
($26.95\pm0.15$ mag in $g_{\rm 475W}$) that lie within 11 kpc of the galaxy centre within the northern half of NGC~1277.
In so doing, we detect $92\pm18$ clusters to our photometric limit, where the uncertainties
come from the uncertainty of $\pm~0.2$ mag in the peak position of the luminosity function.
From this number we then subtracted the expected contribution from NGC~1278 clusters ($27\pm8$ clusters).
Correcting for radial incompleteness, this number becomes
$93.9\pm23.0$ clusters.
We then doubled the total number of clusters to account for the undetected faint half of the cluster
luminosity function, and doubled it again since we constructed the surface density
profile in the northern half of the galaxy. 
We arrive at a total cluster population $N_{\rm cluster} = 376\pm94$ clusters. For $M_V = -20.87$ (NED),
this gives a specific frequency, $S_{\rm N} = 1.7\pm0.4$. This value is consistent with 
galaxies in the stellar mass range of NGC~1277, which typically have $S_{\rm N}\sim 2.0$ \cite{Peng2008}.\\

\subsection*{The blue tilt}

The ``blue-tilt'' manifests as a colour-magnitude relation in the blue clusters such that the brightest
blue clusters become redder with increasing luminosity\cite{Harris2006}\cite{Strader2006}\cite{Mieske2010}.
This can make the colour distributions for the brightest clusters look unimodal.
We investigated this issue by making colour distributions from the median bin values of
  our Monte Carlo simulations for the NGC~1277 clusters binned by magnitude (E.D. Fig. 6).
The mean colours of all three bins are very similar and are predominantly
red with $\langle~g_{\rm 475W}-z_{\rm 850LP}~\rangle = 1.22$ ($22.0 < z_{\rm 850LP} < 24.0$), 
$\langle~g_{\rm 475W}-z_{\rm 850LP}~\rangle = 1.19$ ($24.0 < z_{\rm 850LP} < 25.0$) and 
$\langle~g_{\rm 475W}-z_{\rm 850LP}~\rangle = 1.23$ ($25.0 < z_{\rm 850LP} < 26.5$).
This behaviour is not expected if a blue tilt were driving the observed colour
distributions. The majority of the clusters with $(g_{\rm 475W}-z_{\rm 850LP})_0\leq1.1$
are consistent with the background residuals. However, a few of the brightest
clusters (8) are above the background. These ``blue'' clusters constitute a small fraction of the overall GC population.

\subsection*{The environments of relic galaxies}

NGC~1277 lies 3.8 arcminutes ($\sim80$ kpc) in projection to the North of the central massive
galaxy NGC~1275 in the Perseus cluster of galaxies. Should we be surprised to find a relic galaxy in such
a dense environment?

Observational studies that have looked specifically at the preferred environments
of massive compact galaxies in the nearby universe find that the fraction of present day compact,
quiescent galaxies with masses above $3\times10^{10}$~\msun~ that lie in low density
environments is $\sim4.4$\%~\cite{Poggianti2013}.
The corresponding fraction in nearby galaxy clusters is $\sim22$\%~\cite{Valentinuzzi2010}.
Similar dependencies on environmental density have been obtained
by other studies based on different samples \cite{Peralta2016}\cite{Damjanov2015}.
Of the 16 best-studied nearby massive compact galaxies (including NGC~1277), 7 lie in clusters, 4 in groups and 5
are in isolated systems \cite{Yildirim2017}.

On the theoretical side, \cite{Stringer2015} used DM only simulations to
explore the clustering properties of massive compact galaxies, finding that the
fraction of massive compact galaxies is $\sim5$ times larger in the most massive structures than in low mass DM haloes.
Similarly, \cite{Peralta2016} used semi-analytic models to explore the environments of
compact massive galaxies and found that they represent $\sim0.04$\% of the total galaxy population, but represent $\sim0.18$\%
of the galaxy population in clusters. In addition, \cite{Peralta2016} showed that compact galaxies preferentially
lie in the centres of clusters ($\lesssim0.2$ virial radii -- corresponding to $\lesssim400$ kpc in Perseus)
whereas non-compact galaxies of the same mass have mean clustercentric radii of $\sim0.4$ virial radii.
\cite{Wellons2016} explored the evolution of massive compact systems at $z = 2$ in cosmological, hydrodynamical
simulations. Of their sample of 35 galaxies, $\sim30$\% remained sufficiently undisturbed
as to be defined as relics at $z = 0$. Of these relics, $\sim20$\% survive as satellites
in galaxy clusters.

In summary, although massive, compact ``relic'' galaxies are rare in the nearby Universe
(with space densities $\lesssim10^{-5}$~Mpc$^{-3}$)~\cite{Trujillo2009}\cite{Taylor2010}, they are expected to be preferentially found
in the most massive structures, such as the Perseus cluster of galaxies.

\subsection*{Accretion models for cluster formation}

To get a quantitative understanding of the link between cluster colour distributions and a galaxy's
accretion history, we constructed a library of analytic merger histories for NGC~1277.
We ran $1.5\times10^{6}$ model realizations of NGC~1277 where we reconstruct its total present day
stellar mass by assuming it had accreted some fraction of its
stellar mass ($f_{acc}$) between  $0.0 \leq f_{acc} \leq 1.0$.  In order to test rare, or non-cosmological
merger histories, for each realisation we draw satellite galaxies from a subhalo dark matter mass function with
a randomly selected slope ($-1.0 \leq \alpha \leq 0.0$).  These satellites are stochastically assigned stellar mass via
empirical abundance matching relations \cite{Leauthaud2012} and their stellar mass added to the in-situ
stellar mass of NGC~1277 ($M_{in-situ} = (1-f_{acc}) M_{*,obs}$) until the observed total present day stellar
mass of NGC~1277  ($M_{*,obs}$) is reached for that merger history realization.
  This necessitates a dynamic subhalo mass-function cutoff, in that the maximum accreted subhalo at any point cannot be
  more than the total accreted mass for that trial, and must be less than the in-situ mass.  This ensures a working
  definition of the primary being most massive, and importantly, preserves an accretion fraction of $f_{acc}$ and total
  mass of $M_{*,obs}$ at the end of each trial.
  
The stellar mass of each subhalo
sets its metallicity \cite{Leaman2013}\cite{Kirby2013} (and the metallicity of the clusters which are accreted with it),
and the satellite's dark matter mass provides the specific frequency of clusters for that
satellite \cite{Peng2008}\cite{Georgiev2010} (importantly with stochastic sampling of the scatter
in all the relations).
The same relations are used to construct the cluster population of the in-situ component of NGC~1277 in every realisation.
Finally the cluster metallicities are converted to $g_{475W} - z_{850LP}$ colours~\cite{Peng2006}.

This exercise is qualitatively similar to that employed by \cite{Tonini2013} with the added benefit here
that we allow for a flexible slope of the sub halo mass function, rather than constraining it based on average
accretion histories in CDM simulations.  This allows us to explore how cluster colour distributions and galaxy stellar
mass could be assembled in potentially rare scenarios with either very low or high numbers of mergers or
consecutive extreme mass ratio mergers.

While simple, these models recover the ensemble properties of observed cluster colour and number distributions for galaxies
of the mass of NGC~1277.  E.D. Fig.~7 shows the accreted fraction of each realization of NGC~1277 (with each dot
representing one possible merger history which repoduces the total mass of NGC~1277) versus the number of in-situ to accreted
(red to blue) clusters for NGC~1277 in that realization. Note that we have conservatively used the 
minimum $N_{in-situ} / N_{acc}$ value allowed within the error bars, i.e. $N_{in-situ} / N_{acc} = $2.17 rather
than  $N_{in-situ} / N_{acc} = N_{red} / N_{blue} = 99 / 21 = 4.76.$ 
The observed cluster colour ratio of NGC~1277 clusters is characteristic of
galaxies which have undergone merger histories resulting in the host accreting $\lesssim 12\%$ of the present day
stellar mass.  In comparison, galaxies of comparable mass to NGC~1277 in the ACSVCS sample, show observed colour
distributions ($N_{red} / N_{blue} \leq 1$) typical of having accreted $50-90\%$ of their
stellar mass.
A very small percentage ($\sim 0.02\%$) of these high $f_{acc}$ models can produce cluster colour ratios similar to
NGC~1277.  However, Fig. 4. and E.D. Fig.~8. show that these successful high $f_{acc}$ assembly
histories tend to overpredict the \emph{total} number of clusters per unit mass (i.e., $S_{N} \geq 2$) and tend
to result in systems with DM masses $M_{tot} \geq 10^{14} M_{\odot}$, which is ruled out for NGC~1277 by 
dynamical modelling \cite{Yildirim2017}\cite{Yildirim2015}.

E.D. Fig.~8. summarizes that the typical merger histories in our models which successfully reproduce the $S_{N}$ and
colour distribution of NGC 1277 tend to: have only accreted $f_{acc} = 12 \pm 8\%$ of their stellar mass, had at most a
1:10 mass ratio merger event, and reside in under-massive DM halos ($M_{ *}/M_{DM} \sim 0.06$) when compared to normal galaxies
of the same stellar mass as NGC~1277. These models support the idea that NGC~1277 has not accreted significant amounts
of dark matter rich subhalos leaving it deficient both in DM mass and in blue, accreted clusters.

\subsection*{Code availability}

The photometry software Source Extractor and PSFex are publicly available at https://www.astromatic.net/. The mixture modelling code GMM is publicly available at http://www-personal.umich.edu/~ognedin/gmm/. A version of the GCLF fitting code is available
upon request. The code used for the modelling of the accretion histories of NGC 1277 will be made available in a forthcoming publication.

\subsection*{Data availability}

The HST data used was obtained under GO-14215 (PI: Trujillo) and GO-10546 (PI: Fabian) and is publicly available at https://archive.stsci.edu/. Data products are available upon request.

\section*{Extended Data}

\begin{figure*}
  \begin{center}
\includegraphics[scale=0.8]{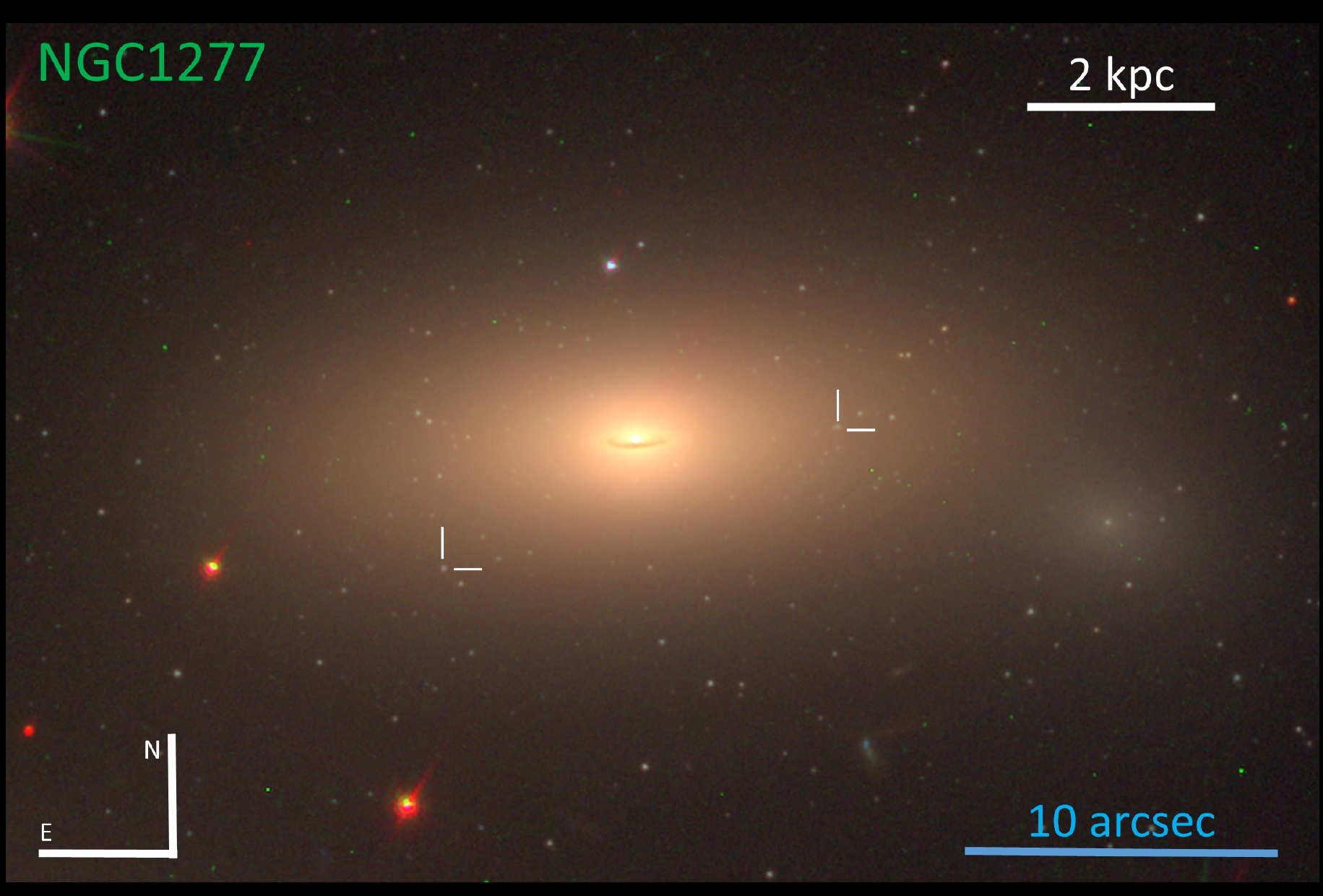}
\caption*{{\bf Extended data Fig. 1. Colour composite ($g_{\rm 475W}$, $r_{\rm 625W}$, $z_{\rm 850LP}$) HST image of the massive relic galaxy
  NGC~1277.} The $g_{\rm 475W}$ and  $z_{\rm 850LP}$ imaging was obtained with the HST programme GO-14215 (PI: Trujillo),
  the $r_{\rm 625W}$ imaging with the programme GO-10546 (PI: Fabian).
  NGC~1277 is the best example found so far in the nearby Universe with characteristics
  equivalent to the first massive galaxies to form more than 11 Gyr ago.
  The image is oriented with North pointing up and East to the left. The field of view
  is $42.2\times30.6$ arcseconds, corresponding to a physical scale of $14.5\times10.5$ kpc at
  our adopted distance of 73.3 Mpc to the galaxy. The images have been scaled with logarithmic intensity
  to highlight the various structures in the galaxy. The vast majority of point sources surrounding
  NGC~1277 are clusters associated with the galaxy. Two example clusters have been marked
  with ticks.
}\label{Fig5}
\end{center}
  \end{figure*}

\begin{figure*}
    \begin{center}
\includegraphics[scale=0.5]{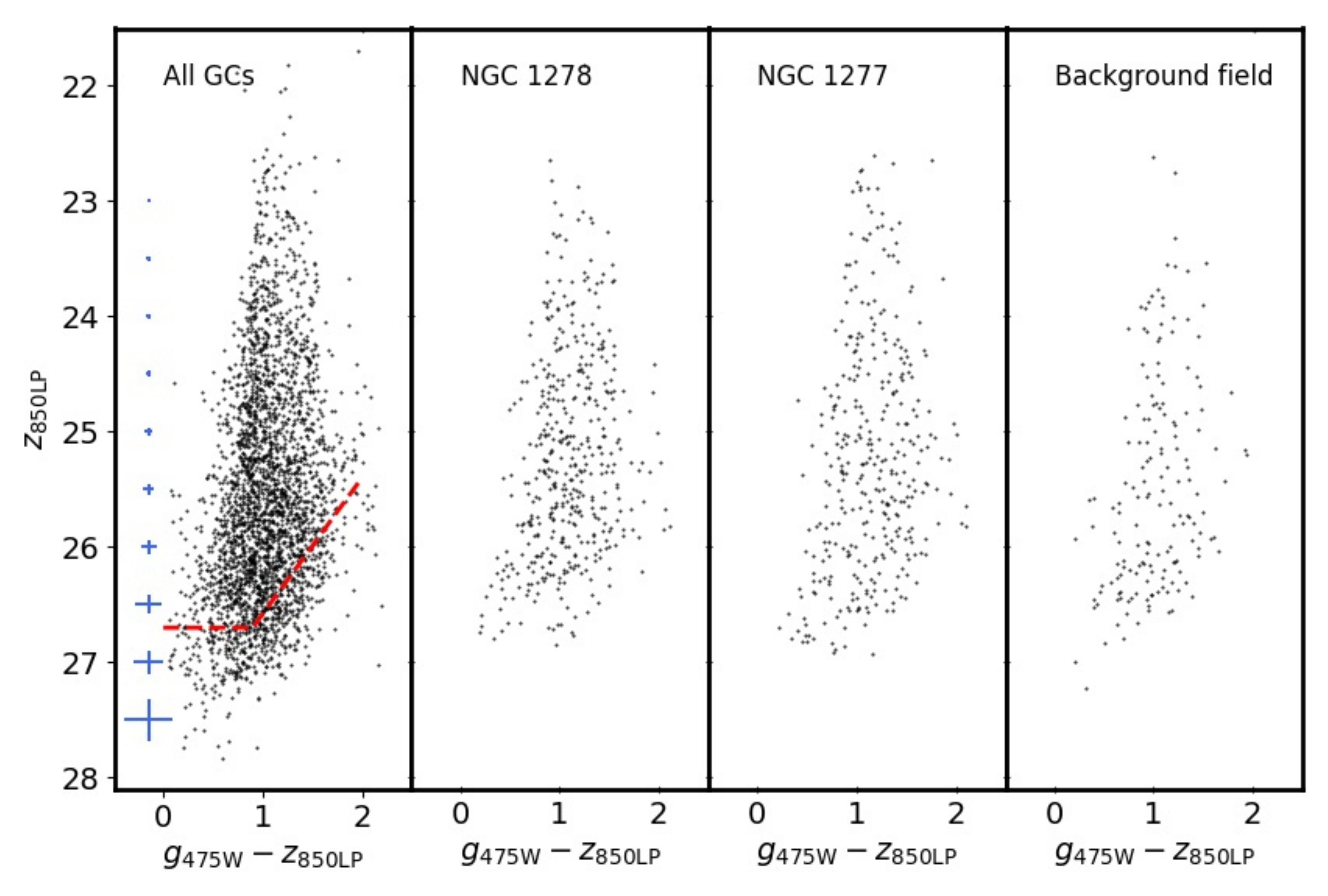}
\caption*{{\bf Extended Data Fig. 2: Colour-magnitude diagrams of cluster candidates.} From left to right:
  all point sources in field, NGC~1278 clusters, raw (not background corrected) NGC~1277 clusters and an example
  background field. Also shown are the photometric uncertainties (1 sigma) based upon
  our artifical point source test, and the 100\% completeness limit (red dashed lines; see Methods).
}\label{Fig6}
    \end{center}
\end{figure*}

\begin{figure*}
    \begin{center}
\includegraphics[scale=0.75]{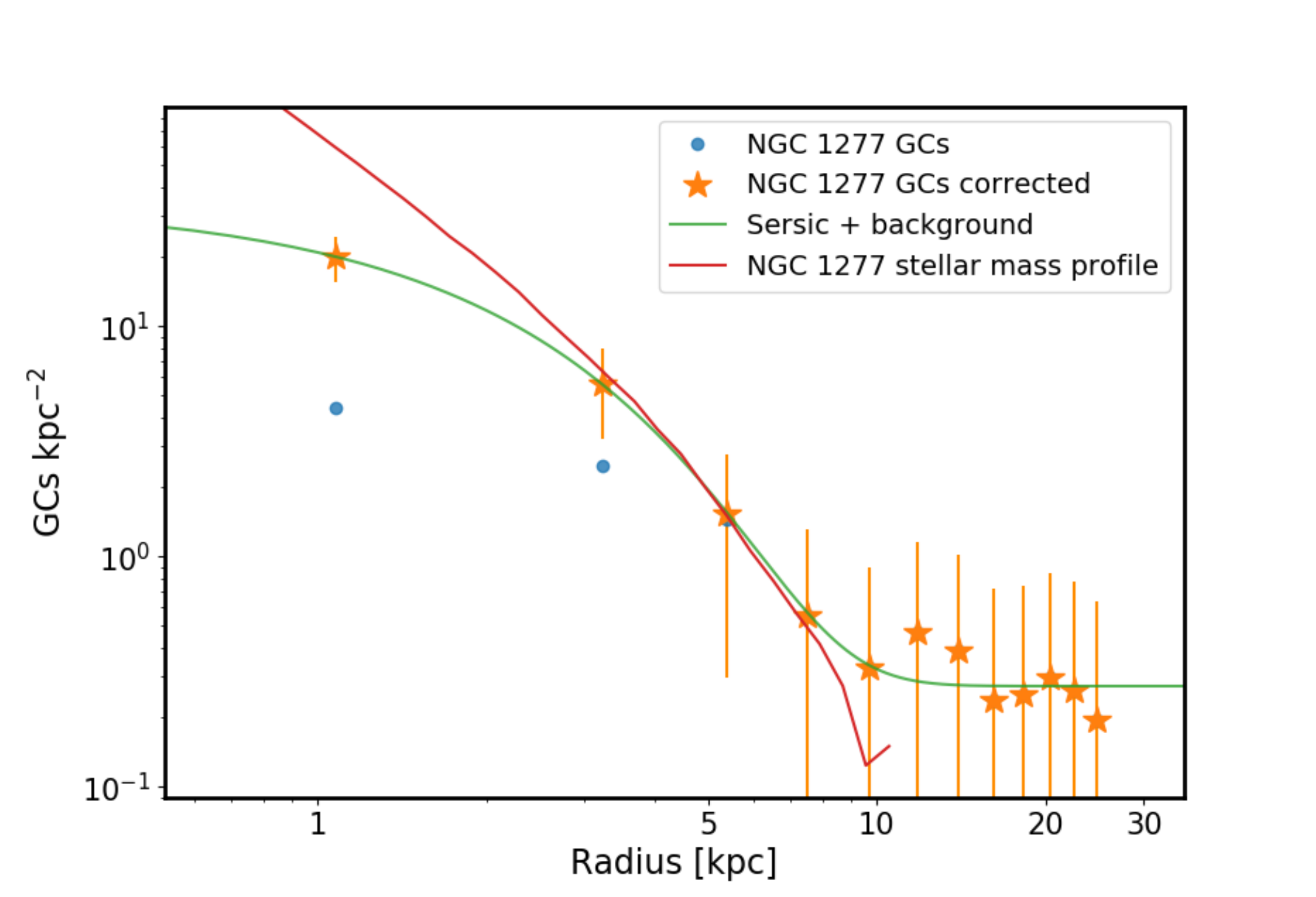}
\caption*{{\bf Extended Data Fig. 3: Surface density profile of NGC~1277 clusters.} 
  The figure shows the raw (blue points) and completeness corrected counts (orange stars).
  Within the range $3-10$ kpc the clusters closely follow the light distribution
  of the galaxy. This is typical of red clusters. The background is determined from
  the S\'ersic function fits, which include a background term, to the cluster data and defines the radial extent of
  the cluster system which ends at 11 kpc ($\sim10$ galaxy $R_e$).
  The stellar mass density profile\cite{Trujillo2014} has arbitrary normalisation.
}\label{Fig7}
    \end{center}
\end{figure*}

\begin{figure*}
    \begin{center}
\includegraphics[scale=0.6]{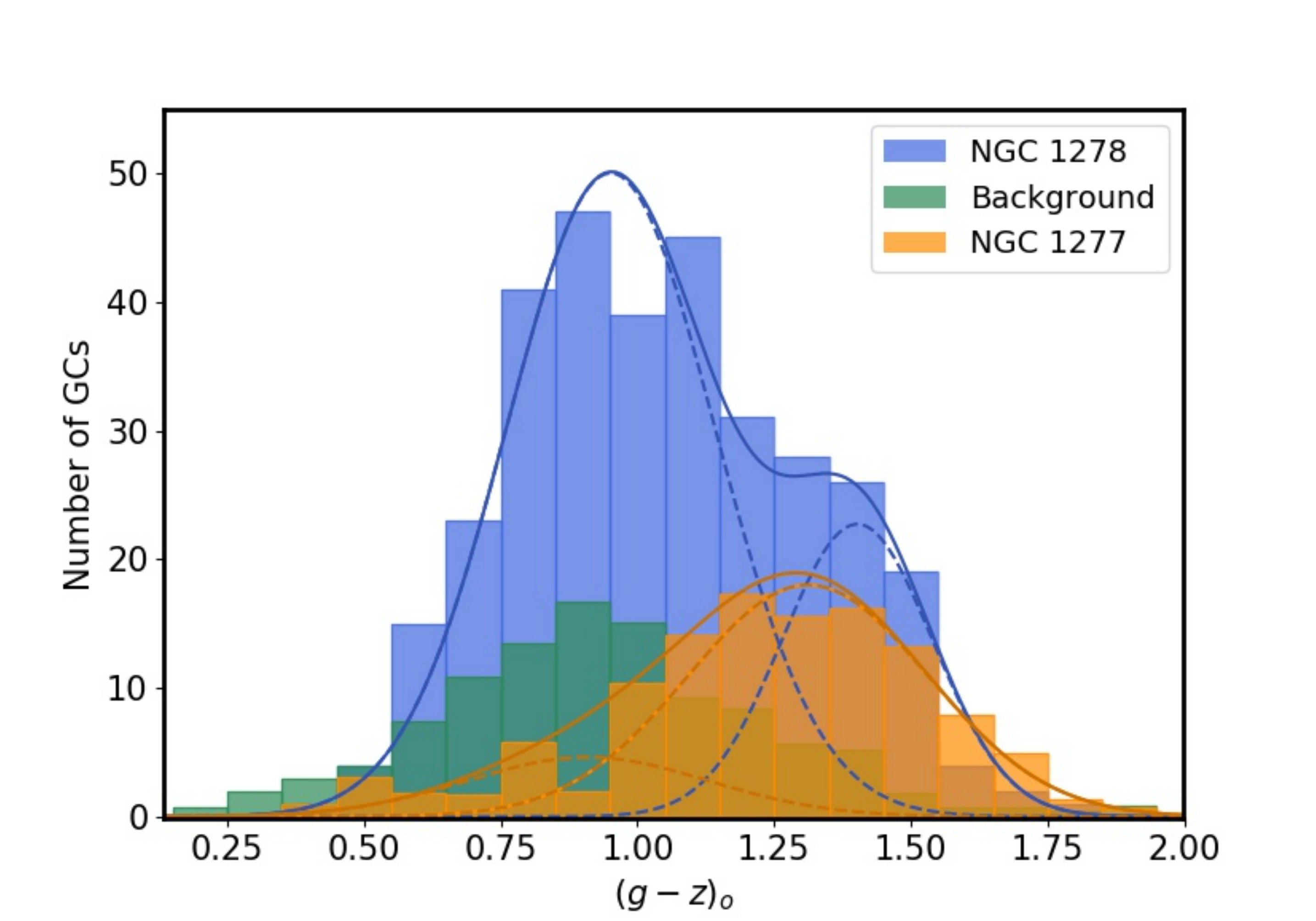}
\caption*{{\bf Extended Data Fig. 4: Colour distribution of NGC~1277 clusters compared to that of the
  companion galaxy (in projection) NGC~1278.} The figure shows that
  NGC~1278 has a strong peak of blue clusters not seen in NGC~1277.
  The expected background contamination of NGC~1278 clusters and intra-cluster clusters to the cluster system of NGC~1277,
  with which we have corrected the colour distribution of NGC~1277, is also shown. The predominantly
  blue colours of the contaminating clusters is as expected since they correspond to the
  outskirts of NGC~1278 (see Fig. 1). Curves are
  two-gaussian solutions from the gaussian mixture modelling code {\sc gmm}\cite{Muratov2010}.
  A single gaussian fit to the NGC~1277 clusters with {\sc gmm} gives
  $\langle~g_{\rm 475W}-z_{\rm 850LP}~\rangle=1.22\pm0.03$ with a
  full-width half-maximum, $0.28\pm0.02$.
}\label{Fig9}
\end{center}
\end{figure*}

\begin{figure*}
    \begin{center}
\includegraphics[scale=0.45]{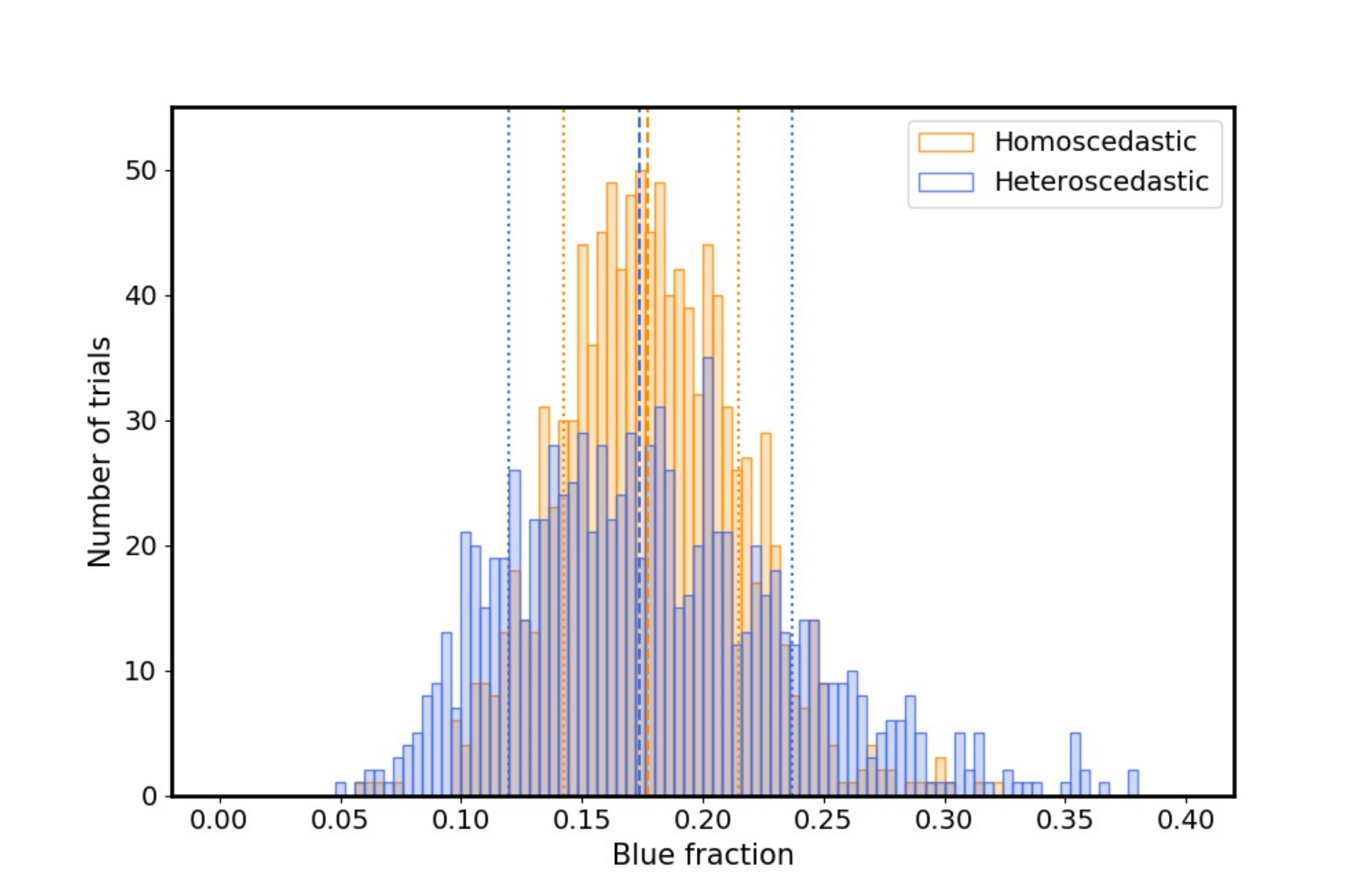}
\caption*{{\bf Extended Data Fig. 5: Distribution of the blue cluster fraction from Monte Carlo simulations.
  The vertical dashed lines indicate the medians of the distributions, and vertical dotted lines show the
  16, 84 percentiles of the distributions. }
}\label{Fig6}
    \end{center}
\end{figure*}

\begin{figure*}
    \begin{center}
\includegraphics[scale=0.45]{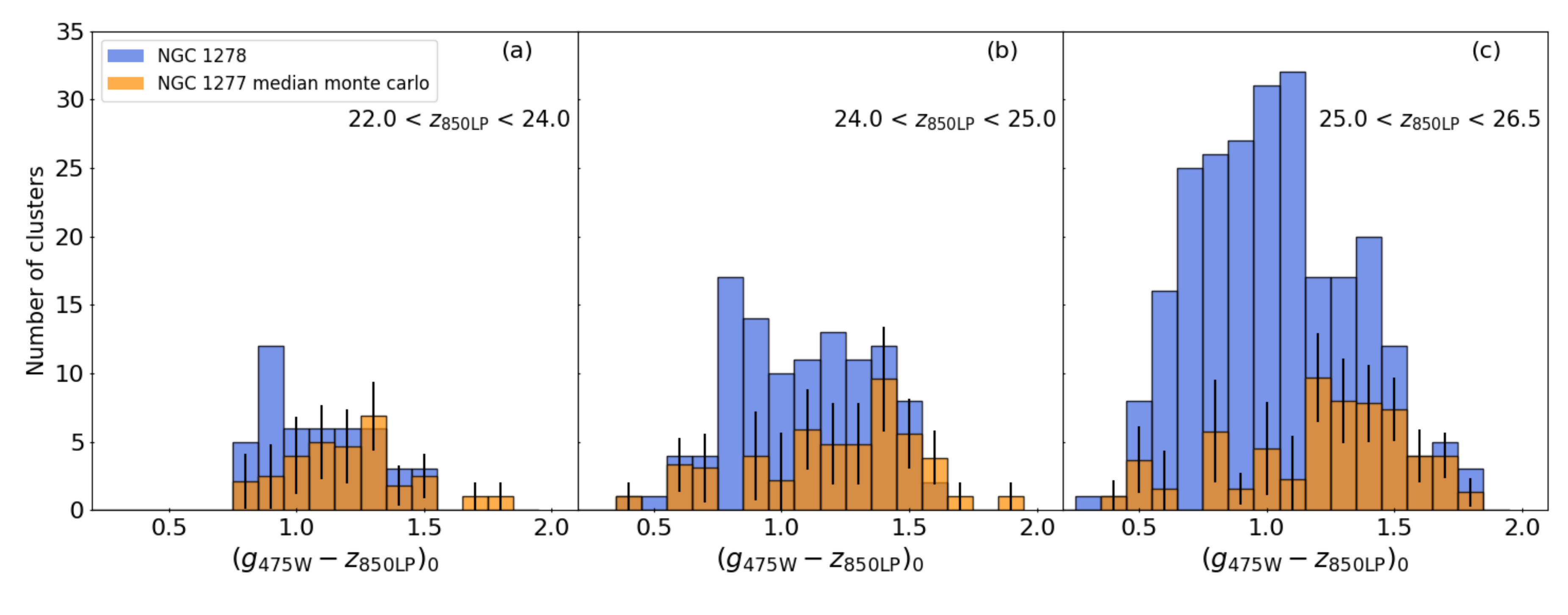}
\caption*{{\bf Extended Data Fig. 6: Colour distributions of NGC~1277 and NGC~1278 in three magnitude bins.
  The bin values for the NGC~1277 clusters represent the medians of the bin values from Monte Carlo simulations.
  Uncertainties are the 16, 84 percentiles of the distributions.  }
}\label{Fig8}
\end{center}
\end{figure*}

\begin{figure*}
    \begin{center}
\includegraphics[scale=0.6]{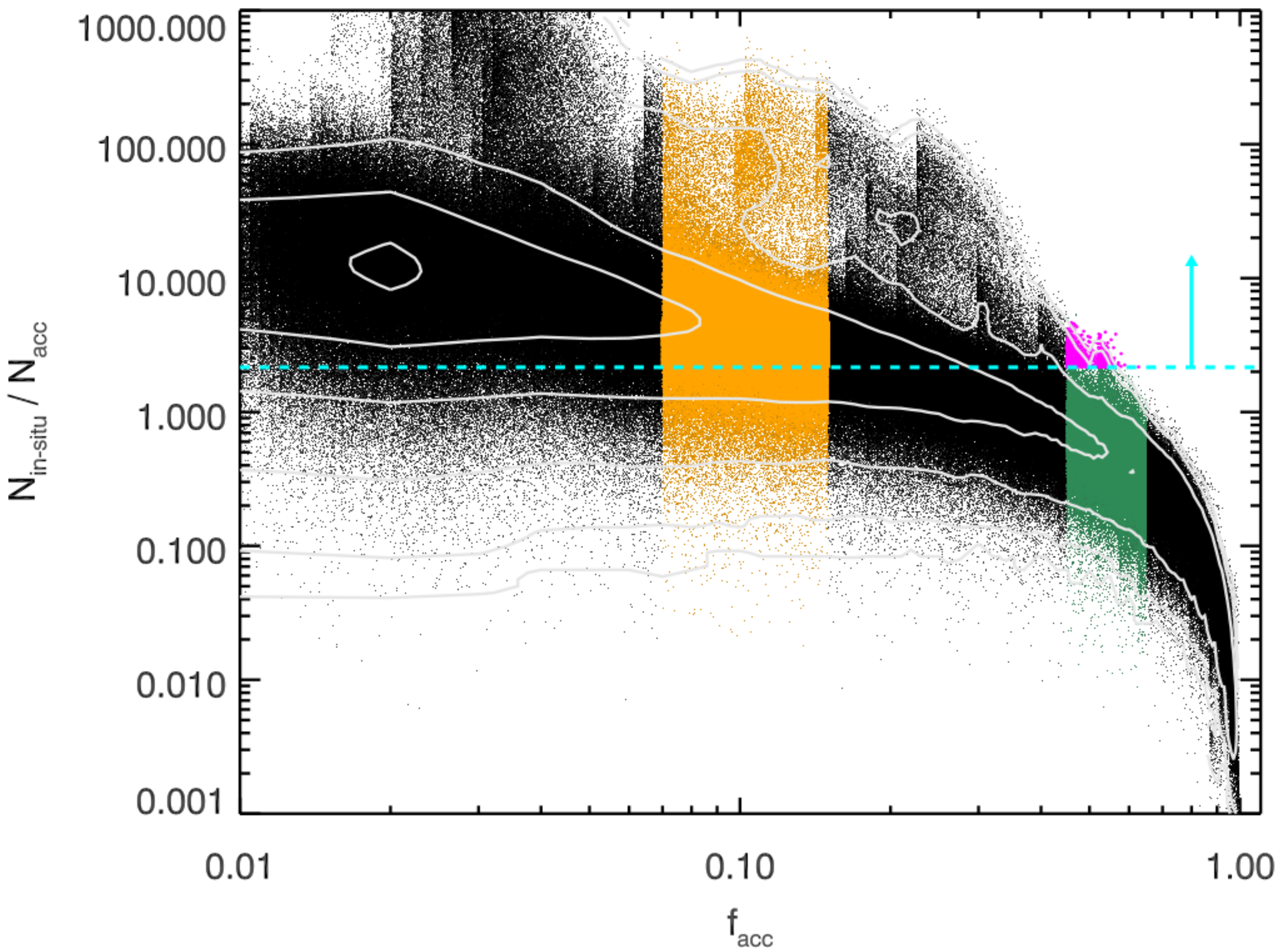}
\caption*{{\bf Extended Data Fig. 7: Accretion models for the build up of the NGC~1277 cluster system.} Each point is a single merger history for NGC~1277, characterized by the total fraction of accreted stellar mass versus the number ratio of in-situ to accreted clusters  for that realization.  The observed ratio for NGC~1277 is indicated by the cyan dashed lines ($N_{blue} / N_{red} =0.21)$, and is characteristic of galaxies with merger histories leading to $f_{acc} \sim10\%$ (orange). Galaxies with comparable stellar mass to NGC~1277 in the ACSVCS, tend to have equal numbers of blue / red clusters, which is more common for accretion histories of $f_{acc} \sim 50-90\%$ (green).
The extremely rare ($\sim0.02\%$) high mass accretion realisations giving the observed blue / red fraction in NGC~1277 are shown in magenta. Contours represent 10, 25, 50, 75, 90 and 99\% of the maximum of the 2D distribution.
}\label{Fig10}
\end{center}
\end{figure*}

\begin{figure*}
    \begin{center}
\includegraphics[scale=0.7]{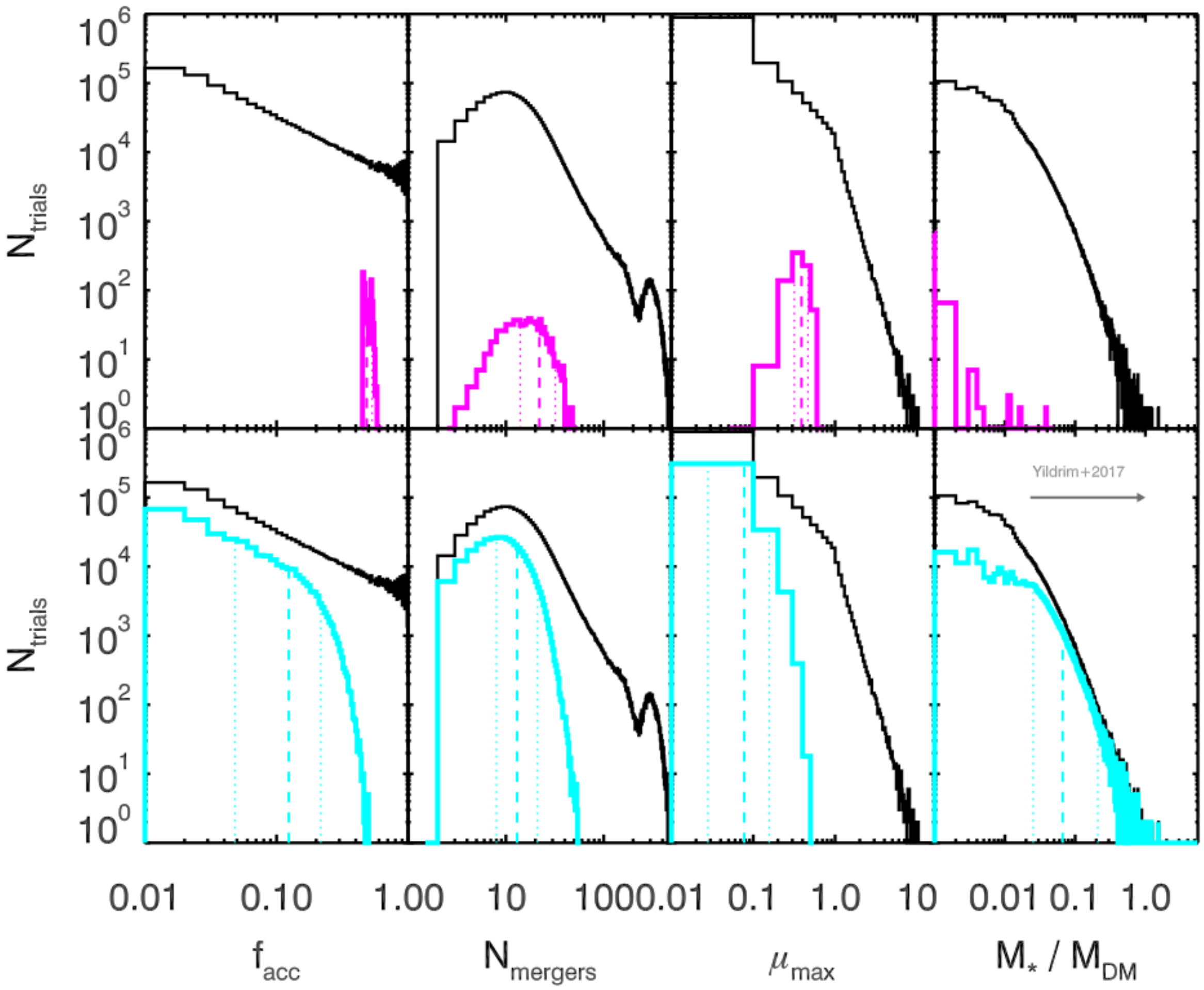}
\caption*{{\bf Extended Data Fig. 8: Accretion fraction, number of mergers, maximum mass ratio of merger, and final stellar to dark matter ratio for all the merger realizations (black distributions).}  The magenta distributions in the top panel show the high $f_{acc}$ merger histories which still have (rarely) satisfied the cluster colour ratio constraint (e.g., E.D. Fig.~7).  The cyan distributions in the bottom panels show the low $f_{acc}$ merger histories which have satisfied the observed cluster colour ratio and $S_{N}$ constraints.  
  Notably, the high accretion fraction models (magenta) also predict a DM halo mass for NGC~1277 that is larger than the observational constraints\cite{Yildirim2017}.​ Vertical dashed lines represent the median of the distributions, vertical dotted
  lines show the 16th and 84th percentiles of the distributions.
}\label{Fig11}
\end{center}
\end{figure*}

\newpage

\section*{References}

\begingroup
\renewcommand{\section}[2]{}%

\endgroup

\end{document}